\documentclass[prd, reprint, superscriptaddress, preprintnumbers, nofootinbib, amsmath, amssymb, aps]{revtex4-1}
\pdfoutput=1
\usepackage{hyperref}
\usepackage{graphicx}
\usepackage{makecell}
\usepackage{subfigure}
\usepackage[textwidth=1.25cm,textsize=footnotesize]{todonotes}
\usepackage[utf8]{inputenc}
\usepackage{color}
\usepackage[normalem]{ulem}

\renewcommand{\l}{\left(}
\renewcommand{\r}{\right)}

\begin{document}
\title{Constraints on the parameters of keV-scale mass annihilating Dark Matter obtained with SRG/ART-XC observations}

\author{E.I. Zakharov}
\email[]{ezakharov@cosmos.ru}
\affiliation{Space Research Institute of the Russian Academy of Sciences, Moscow 117997, Russia}
\affiliation{National Research University Higher School of Economics, Moscow 101000, Russia}
\affiliation{Institute for Nuclear Research of the Russian Academy of Sciences, Moscow 117312, Russia}
  
\author{V.V. Barinov}
\affiliation{Institute for Nuclear Research of the Russian Academy of Sciences, Moscow 117312, Russia}
 
\author{R.A. Burenin}
\affiliation{Space Research Institute of the Russian Academy of Sciences, Moscow 117997, Russia}
\affiliation{National Research University Higher School of Economics, Moscow 101000, Russia}

\author{D.S. Gorbunov}
\affiliation{Institute for Nuclear Research of the Russian Academy of Sciences, Moscow 117312, Russia}
\affiliation{ Moscow Institute of Physics and Technology, Dolgoprudny 141700, Russia}

\author{R.A. Krivonos}
\affiliation{Space Research Institute of the Russian Academy of Sciences, Moscow 117997, Russia}
\affiliation{Institute for Nuclear Research of the Russian Academy of Sciences, Moscow 117312, Russia}

\bigskip
\preprint{INR-TH-2024-0XX}

\begin{abstract}
In this paper we present new constraints on the velocity-independent cross section of keV-scale mass annihilating Dark Matter particles obtained with SRG/ART-XC after 4 full-sky surveys. These constraints are derived from observations of the Milky Way Halo, 33 Local Group spheroidal dwarf (dSph) galaxies and separately for the dSph galaxy Ursa Major III/UNIONS 1. The constraints from the Milky Way Halo are the strongest among others and among all available in literature for this class of Dark Matter models with particle masses from 4 to 15 keV. 
\end{abstract}

\maketitle

\section{Introduction}
A stability of Dark Matter (DM) particles (on cosmological timescales) is typically attributed to a specific symmetry. The simplest example is $Z_2$ (e.g. $R$-parity in supersymmetric models), so the Lagrangian of the model is invariant upon the change of the sign in front of the DM field whatever it is. Generically in this case a possibility remains for two DM particles to annihilate when meet and hence one hopes to search for a product of this annihilation wherever it happens. In more complicated models like $Z_n$, $n>2$, symmetries, a many body initial state can annihilate, and still one can hope to find a hint of this event, if the final state includes some known particles which properties within the Standard Model of particle physics (SM) are well studied. 

The annihilation may proceed in the galaxies which masses are always dominated by nonrelativistic DM particles. The simplest case is when only two particles are present in the final state. As to the signature, it matters not what symmetry is responsible for the DM stability and how many DM particles come in the initial state, since all of them are confined in a galaxy, i.e. are sufficiently nonrelativistic, and hence the energy of outgoing SM particles are fixed by the kinematics. It provides with a peaklike signature in their cosmic spectra, if no significant attenuation for the SM particles in the cosmic space on their way to the Earth is expected.

This is the situation for annihilation into a pair of photons in the keV range, which is the relevant signature for models with DM particles of masses in the same range, for particular models see, e.g.,\cite{Goudelis:2018xqi,DEramo:2020gpr,Cheek:2024fyc}. Absence of any direct evidence of such light particles is natural if they only feebly couple to the visible sector, for reviews see, e.g., \cite{Lanfranchi:2020crw,Agrawal:2021dbo}. The searches of the peak, which frequency is determined by half of the total mass of annihilating DM particles, that is just the mass of DM particle,
\[
\omega_\gamma=m_\chi\,,
\]
in case of $Z_2$ symmetry and $2\to2$ annihilation, 
\begin{equation}
\label{annihilation}
\chi+\chi \to \gamma+\gamma\,,
\end{equation}
have been performed previously with analysis of data collected by $X$-ray telescopes while observing the Milky Way (MW), M31, dwarf galaxies and other potential astrophysical sources, see Refs.\,\cite{nustarM31,Laha:2020ivk,Siegert:2021upf,Siegert:2024hmr, Jeltema:2008ax,Planck}.  

Here we report on the similar study based on 4 full-sky $X$-ray surveys by SRG/ART-XC telescope. The obtained results, which we present as an upper bound on the DM annihilation cross section into a pair of photons \eqref{annihilation} 
are the strongest in the DM mass range $m_\chi$ from 4 to 15\,keV. Our limits are applicable also to the models with annihilating processes $\chi\chi\to \chi+\gamma$ and to annihilation into photon and neutrino, $\chi\chi\to \gamma+\nu$. In these cases the photon frequencies are related to the DM particle mass $m_\chi$ as $\omega_\gamma=3\,m_\chi/4$ and $\omega_\gamma=m_\chi$, correspondingly.

\section{SRG observatory}
The Spektr-RG (Spectrum-Roentgen-Gamma or SRG) is an orbital astrophysical observatory launched into a halo-orbit around the Lagrangian point L2 of the Earth-Sun system on July 13, 2019\,\cite{SRG}. The observatory's primary mission is to scan the entire sky and build a map of the Universe in soft and hard X-rays. The main instruments of the observatory are two X-ray telescopes: eROSITA, operating in soft (0.3$-$10~keV) energy band \cite{eROSITA} and \textit{Mikhail Pavlinsky} ART-XC, operating in hard (4$-$30~keV) energy band \cite{ARTXC}. Both telescopes are equipped with grazing incidence optics based on the Wolter type I scheme.

The main operating mode of the SRG observatory is a ``survey mode''. In this mode, the observatory is rotating around an axis close to the Sun's direction with a period of 4 hours, with the rotation axis shifting by approximately 1$^\circ$ per day following the motion of the Sun. Thus each of the observatory's telescopes covers the entire sky in approximately 6 months. The adopted strategy of the SRG survey leads to the appearance of ``deep fields'' around the North and South Ecliptic Poles, where the big circles of all individual scans cross. As a result, the exposure time depends on ecliptic latitude $\theta$ as $1/\cos(\theta)$, i.e. exposure is minimal at the ecliptic equator and maximal at the ecliptic poles.

\section{ART-XC telescope}
The \textit{Mikhail Pavlinsky} ART-XC (Astronomical Roentgen Telescope -- X-ray Concentrator) telescope \cite{ARTXC, 2018ExA....45..315P, 2019ExA....47....1P, grasp} is one of two X-ray telescopes at the SRG observatory. The telescope consists of seven co-aligned ``mirror system + detector'' pairs. Each mirror system is  designed in Wolter type I scheme with primary paraboloid and secondary hyperboloid mirrors and contains 28 nested shells. Each shell is coated with a thin layer of iridium, which allows to enhance high-energy reflectivity up to 30~keV. The incoming X-rays photons, reflected first by parabolic and second by hyperbolic surface, produce direct X-ray sky image with field-of-view (FOV) of ${\sim}36'$ in diameter (or ${\sim}0.3$ deg$^2$). The angular resolution of the optics with such double reflected photons is about $53''$ \citep[full width at half-maximum;][]{ARTXC}. The photons, reflected only once,  can also be registered with the SRG/ART-XC in a so-called ``concentrator mode'', however, without image reconstruction \citep{grasp}. Such single reflected photons can reach detector from angular scales up to ${\sim}50'$, subtending FOV of ${\sim2}$~deg$^2$. The SRG/ART-XC detectors are double-sided strip detectors (DSSD) made of CdTe single crystal 30x30x1 mm with 48 strips on each layer. Registered event energies cover a range from 3 to 120 keV.

In addition to recorded energy of the event, a characteristic parameter such as the grade of the event is also important for  this study. Events (induced by a photon or a  charged particle) that only trigger a single strip in each layer of the detector have a grade value of 0. Events that trigger both layers, but in one or both layers 2 neighbor strips are triggered, have grade values of 1--8. Events that trigger 3 strips in one or both layers have grade values of 9--15. The vast majority of 9--15 grade events probably originate from interactions of charged particles with the detector and should be excluded from science analysis. In the first part of our work we use only 0 grade events, since the probability that they are triggered by incoming photons is slightly higher compared to 0--8 grades usually used for scientific analysis. In the second part, to increase the event statistics, we use events with grades from 0 to 8.

\begin{figure}
    \centering
    \includegraphics[width=1\linewidth]{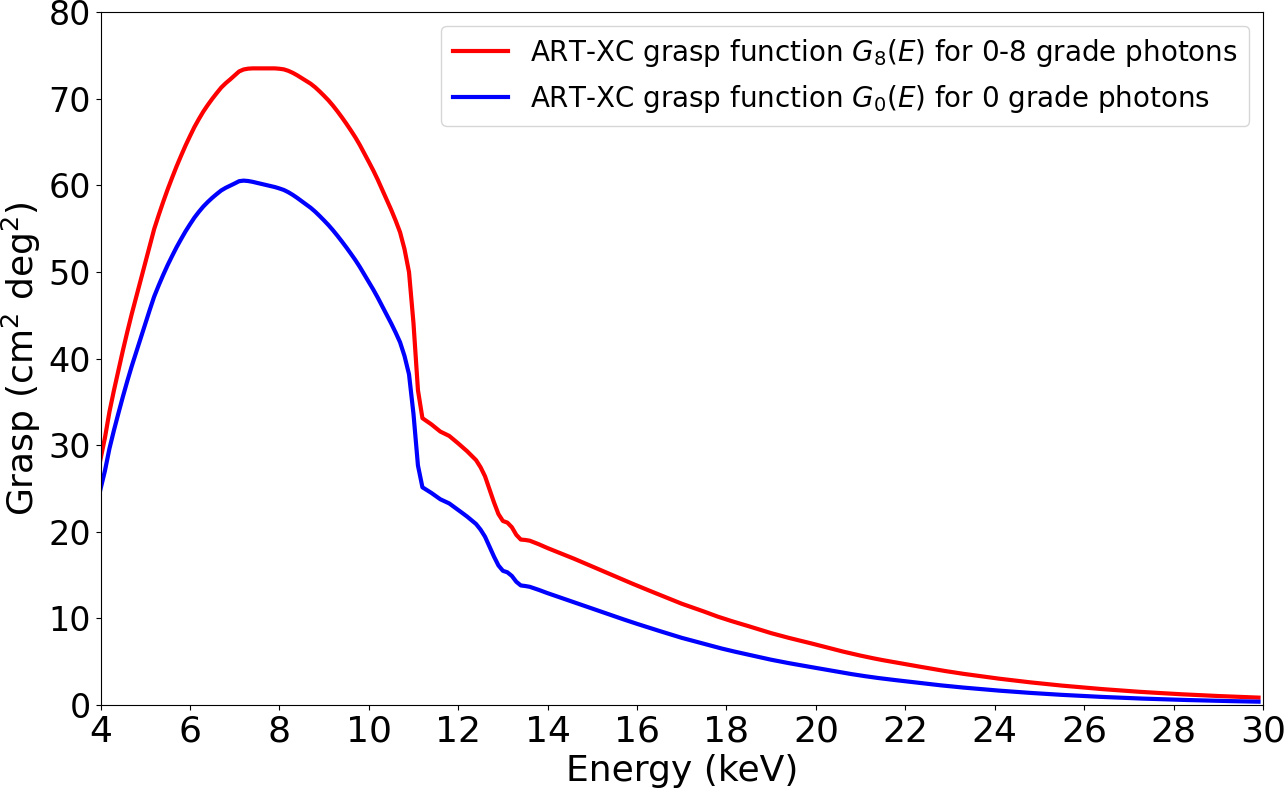}
    \caption{The SRG/ART-XC grasp functions $G_0(E)$ and $G_8(E)$ in concentrator mode for zero grade events only (blue line) and 0-8 grade events (red line), respectively. In 4-30 keV energy range the ratio is $G_0(E)/G_8(E) \approx 0.94 - 0.2(E/\text{keV})$.}
    \label{fig:grasp}
\end{figure}

When observing areas of the sky with an angular size exceeding the FOV of the telescope, the grasp function becomes an important characteristic \cite{grasp}. The grasp is the product of the effective area of the telescope (in cm$^2$) and the FOV (in deg$^2$). Fig.\,\ref{fig:grasp} shows the ART-XC grasp function in the concentrator mode for events with grade 0 only and the combined grades 0--8.

\section{Milky Way halo}
In this part of the current study, we closely follow the analysis described in our previous work \cite{artxc_sndm}. The key point is that the expected signal from the MW Halo is not uniformly distributed across the sky, unlike the Cosmic X-ray Background (CXB). The instrumental background is also independent of the spacecraft orientation in space and is very stable. For more details, see below.

\subsection{ART-XC all-sky data}
The focused X-ray photons that passed through the telescope's mirror system and hit the detectors can be divided into 4 components. The first component is photons from bright point and extended sources. At the same time, galactic sources are concentrated in the sky near the Galactic Plane, and extragalactic sources are uniformly distributed across the sky. The second component is the CXB, which is a diffuse radiation uniformly distributed across the sky. The third component is Galactic ridge X-ray emission \cite[GRXE, ][]{GRXE}, an extended X-ray emission with low surface brightness observed from direction toward the Galactic Plane. The fourth component, the existence of which we assume in our work, is photons from annihilating DM. This radiation is associated with the Galactic Halo and it is distributed inhomogeneously across the sky. The density of this radiation decreases with distance from the Galactic Center (GC).

In addition to focused X-rays, instrumental background must also be considered. The instrumental background is produced by the interaction of solar wind protons and Galactic cosmic rays with the entire structure of spacecraft. This interaction creates an induced radiation, which is registered by the telescope's detectors. The instrumental background does not depend on the orientation of the spacecraft in space and is very stable on the scale of a few hours (one turn of the spacecraft around its axis)\footnote{Space weather monitor at Earth-Sun Lagrange point L2: \url{https://monitor.srg.cosmos.ru/}}. The most significant influence on the instrumental background is exerted by the increasing solar activity. Due to a steady increase of the solar activity, the instrumental background was steadily decreasing during the first 2 years of observations with relative change of about 10\%. For SRG/ART-XC, the ratio of CXB to instrumental background is about 4\%.

Based on all this, we divided the entire sky into the two regions. The first region (Region I or "signal + background") is a cone with half apex angle of $60^\circ$, directed to the GC. As the second region (Region II or "background") we took the rest of the sky. All 1545 point and extended sources detected by the telescope over 2 years in the survey mode (ARTSS 1-5 catalog \cite{ARTSS15}) were masked out with  a radius of $1^\circ$. We also additionally cut out the Coma and Virgo clusters, as they were not originally included in the ARTSS 1-5 catalog. The cutout radius for these sources is $3^\circ$ and $2.5^\circ$, respectively. To remove the influence of GRXE, we cut out a strip of sky with galactic latitude $|b|<1^\circ$. 

\begin{figure}
    \centering
    \includegraphics[width=\columnwidth]{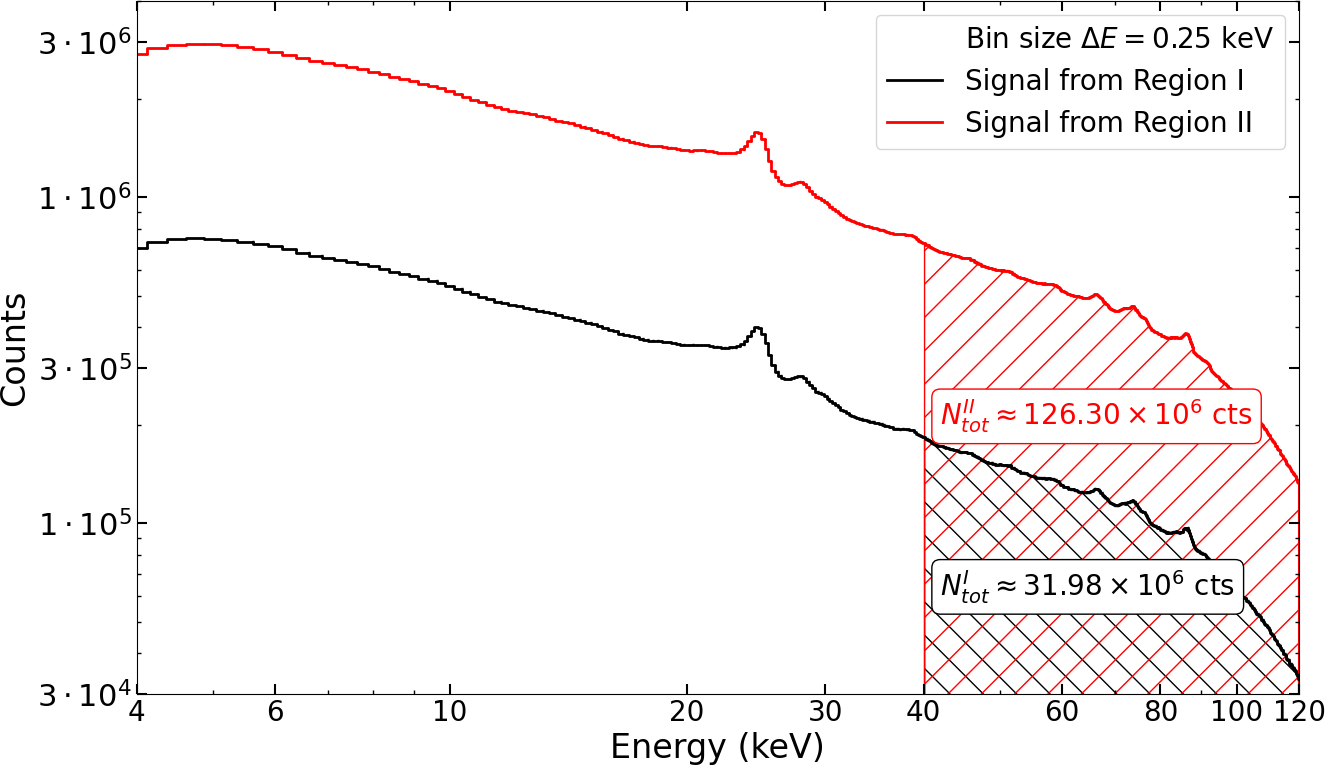}
    \caption{SRG/ART-XC 0-grade events raw spectrum in counts for Region I (black line) and Region II (red line).}
    \label{fig:total_signal}
\end{figure}

\begin{figure}
    \centering
    \includegraphics[width=\columnwidth]{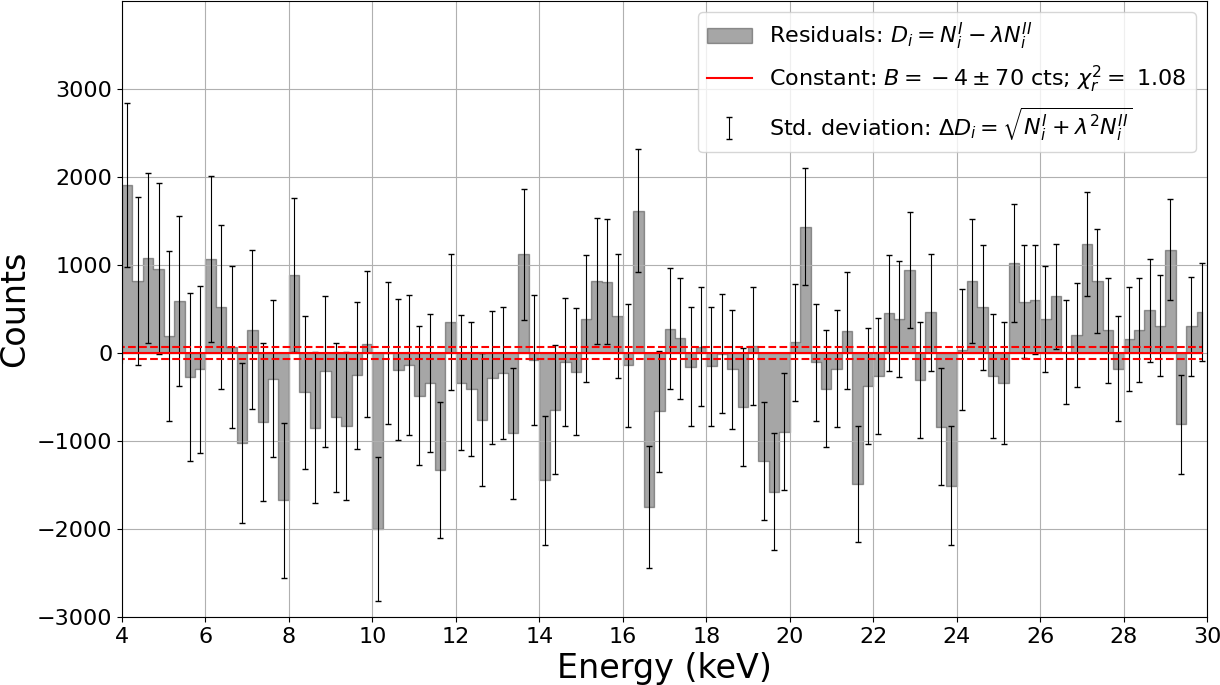}
    \caption{The residuals $D_i$ shown in gray shading, vertical lines show the error bars $\Delta D_i$ \eqref{eq:residuals}. The red solid line is the fit to the residuals by constant $B$ and the red dashed lines indicate $\pm1\sigma$ confidence intervals.}
    \label{fig:residuals}
\end{figure}

Fig.\,~\ref{fig:total_signal} shows the raw spectra from Regions I and II. Since all background components are uniformly distributed across the sky, while the useful signal is not, the difference in the spectra between Regions I and II allows to find the signal of interest. To subtract the spectra, we use a cross-normalization factor between sky regions as the total  number of detected events in 40$-$120~keV band,  $\lambda = N^I_{\rm tot}/N^{II}_{\rm tot} = \omega^I T^I/\omega^{II} T^{II} \approx T^I/T^{II} = 0.25321 \pm 0.00005$, where $\omega$ and $T$ are the count rate and exposure time in the corresponding regions.  Fig.\,\ref{fig:residuals} shows the residuals calculated applying the expression

\begin{equation}
    \label{eq:residuals}
    D_i \pm \Delta D_i = \l N^I_i - \lambda N^{II}_i \r \pm \sqrt{N^I_i+\lambda^2 N^{II}_i}.
\end{equation}

The residuals are characterized by a weighted mean $B = -4 \pm 70$~cts with reduced $\chi^2_{\rm r} = 1.08$. This value indicates that the shapes of the spectra in both sky regions are similar to each other.

\subsection{Dark Matter signal} 
\begin{figure}[tb]
    \centering
    \includegraphics[width=\columnwidth]{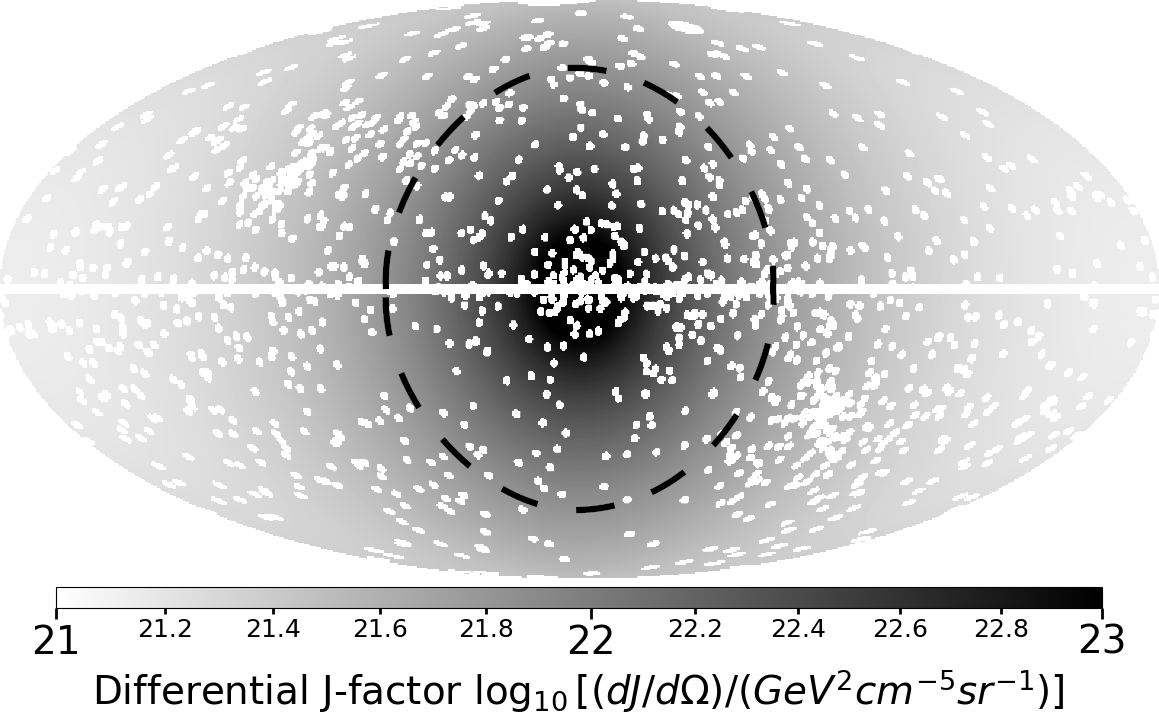}
    \caption{Differential J-factor for NFW profile \cite{McMillan2017} of the MW Halo. The regions excluded from our analysis are shown in white (excluded area $\approx 4700$ deg$^2$). Black dashed circle denotes the border between the Region I (inner area, $\Omega^I \approx 9000$ deg$^2$) and Region II (outer area, $\Omega^{II}\approx 27500$ deg$^2$). Regions I and II comprise exposure time of $T^I \approx 10.4$\,Ms and $T^{II}\approx 41.4$\,Ms, respectively (for exposure map see \cite{artxc_sndm}). Integral J-factor for Region I obeys $ 22.85=\log_{10}(J^I/(\text{GeV}^2\text{cm}^{-5})$); for Region II we find $22.37=\log_{10}(J^{II}/(\text{GeV}^2 \text{cm}^{-5})$).}
    \label{fig:DM}
\end{figure}

The differential flux from annihilating DM in a given direction is determined by its J-factor $dJ=d\Omega\int\ \text{d} l\, \rho^2_{DM}(\vec l)$  along the line of sight.  Hence the signal flux expected from the central part of the MW equals
\begin{equation}
    \label{eq:flux}
    F^I = \frac{\sigma v}{4 \pi m^2_\chi} \times J^I
\end{equation}
where the J-factor is
\begin{equation}
    \label{eq:Jfactor}
    J^I = \int_{0}^{2\pi}
    \!\!\!
    \int_0^{60^\circ}
    \!\!\!\!
    \int_0^{R_\text{vir}}
    \!\!\!
    \rho^2_{\text{DM}} (r(l,\theta)) \sin(\theta) \text{d}\phi \text{d}\theta \text{d}l
\end{equation}
Here the distance from the MW center is $r(l, \theta) = \sqrt{R^2 - 2R l\cos\theta + l^2} $, $\theta$ is the angle from the direction to the MW center, $ R=8$\,kpc is the distance from the observer to the MW center, the virial radius is $R_\text{vir}=200$\,kpc \cite{Dehnen:2006cm} and $l$ is the distance along the line of sight. The signal flux from the second region, $F^{II}$ is given by \eqref{eq:Jfactor} upon replacing the integration range $(0,60^\circ)\to (60^\circ,180^\circ)$. As a main DM profile in this analysis we use the standard Navarro-Frenk-White (NFW) profile $\rho_{\text{\tiny DM}}(r) = \rho_{s}/\left(r/r_s\right)\left(1 + r/r_s\right)^{2}$ \cite{NFW} with $\rho_{s} = 8.54\times 10^{-3}\text{M}_{\bigodot} \text{pc}^{-3}$ and $r_s = 19.6$\,kpc~\cite{McMillan2017}. Fig.\,\ref{fig:DM} illustrates the distribution of the J-factor across the sky.

\subsection{Calculating the limits} \label{CalcLim}
The spectrum of photons emitted at annihilation is highly monochromatic and is well described as $dN/dE_{\gamma \gamma} \approx 2\delta(E-m_\chi)$. However, after interaction with the detector, the photon spectrum is a convolution of the original spectrum and the instrumental energy response. We approximate SRG/ART-XC instrumental response with the Gaussian characterized by full width at half maximum (FWHM) as a function of energy presented in Ref.\,\cite{ARTXC}. It is almost constant, FWHM($E$)$\approx 1.2$\,keV with deviations not exceeding 10\%, and hence for the number of signal events $S_i$ in the energy band from $E_i$ to $E_i+\Delta E$, where $\Delta E=0.25$\,keV, one gets 
\begin{equation}
    \label{eq:gauss}
    S_i=\frac{A\times 2\sqrt{\log 2}}{\sqrt{\pi} \text{FWHM}(E)} \exp\l-\l\frac{ 2(E_i-E)}{\text{FWHM}(E)}\r^2\log{2}\r \Delta E,
\end{equation}
with parameter $A$ related to the signal flux as
\begin{equation}
    \frac{A\Omega^I}{T^I G_0(E)}=F^I-\lambda F^{II}.
\end{equation}
To constrain $A$ at a given energy $E$ (and hence the velocity-independent cross section $\sigma v$ at a given mass $m_\chi$) we construct the log-likelihood function 
\begin{equation}
    \label{eq:LogLike}
    l(D|A,E) = \sum_{i=1}^{N^b} \frac{(D_i-B-S_i(A,E))^2}{2\sigma_i^2},
\end{equation}
where $\sigma_i^2=\l\Delta D_i\r^2$, i.e. the main errors are pure statistical. We use a Gaussian likelihood function because of the huge number of events registered by the telescope detectors. It can be seen from Fig.\,\ref{fig:total_signal} that in both Regions each bin contains millions of events and therefore has a Gaussian distribution. So, their difference $D$ in each bin will also have a Gaussian distribution. This fact distinguishes our case from the classical "On-Off measurement" described in \cite{LiMa}\footnote{We also note that all three approaches to signal significance estimation given in \cite{LiMa} are fully equivalent in the case of Gaussian statistics.}.

\begin{figure}[bt]
    \centering
    \includegraphics[width=\columnwidth]{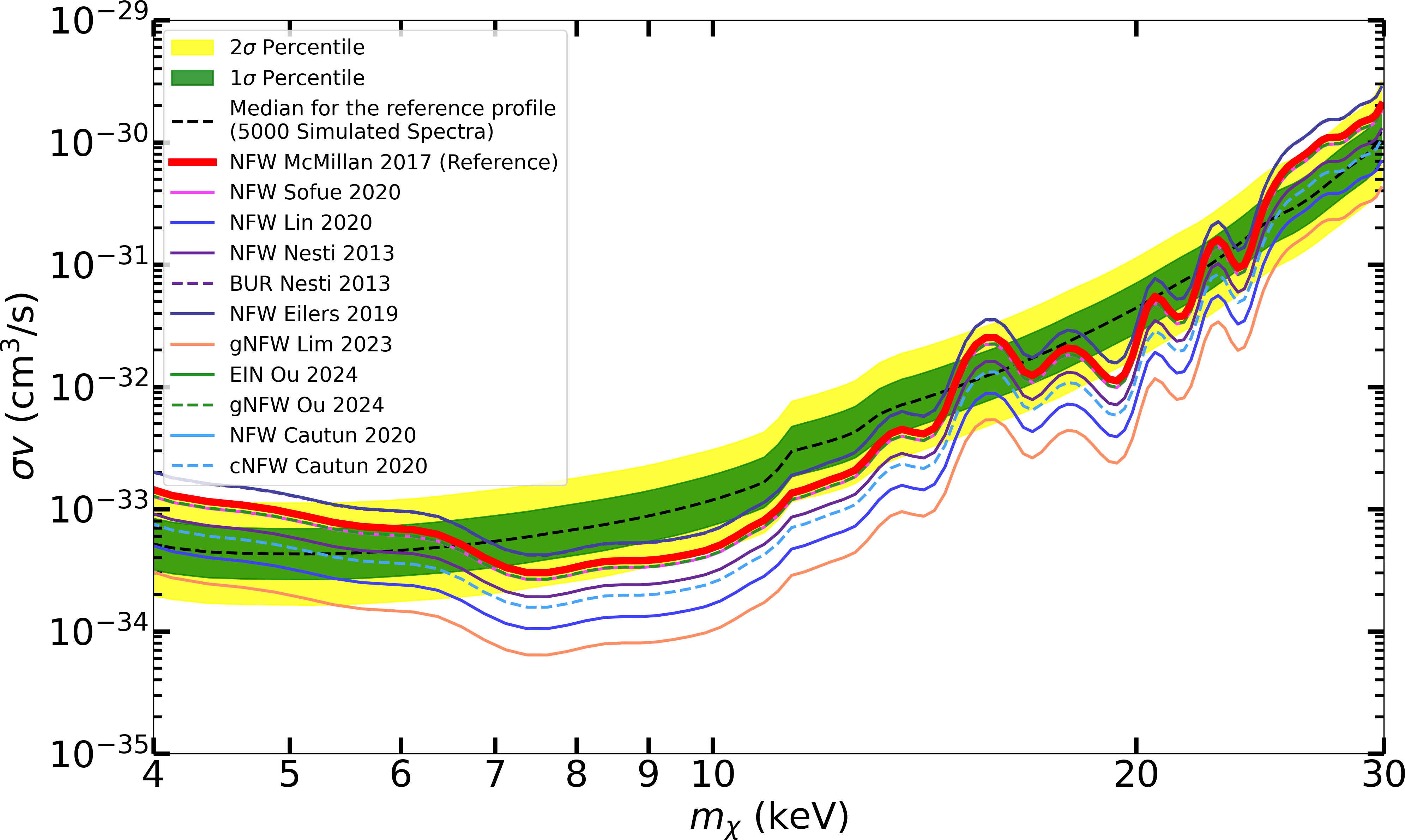}
    \caption{Upper limits on s-wave annihilation cross section (95\% C.L.) obtained with SRG/ART-XC in the MW Halo after 4 full-sky surveys. The red solid line shows the constraints for the reference DM profile \cite{McMillan2017}. The black dashed line shows the median value of the expected constraints obtained with analysis of 5000 generated spectra for the reference DM profile. The green and yellow areas correspond to 1$\sigma$ and 2$\sigma$ percentiles, respectively. Other colored lines show the constraints for various DM profiles \cite{Eilers2019, Sofue2020, Ou2024, Cautun2020, Lim2023, Nesti2013, Lin2019}.}
    \label{fig:resultsMWHalo}
\end{figure}

To find the upper limit for area under the line $A$, we perform the following procedure. For each energy we increase $A$ to make the following equality true 
\begin{equation}
    \label{eq:95cl}
    2 \times \l l(D|A,E) - l(D|\hat{A},E)\r = 2.71,
\end{equation}
where $\hat{A}$ is the maximum likelihood estimate of A. In the large count limit, the log-likelihood difference reduces to $\Delta \chi^2$ for a single degree of freedom. In terms of Gaussian standard deviations, one side 95\% C.L. upper limit corresponds to $1.64\sigma$, i.e.\ to $\Delta \chi^2 = 2.71$. Consequently we constrain the product of annihilation cross section and the DM relative velocity $\sigma v$ treating it as velocity-independent,  which is true in case of the s-wave annihilation. Thus obtained limits depend on the DM mass, the results are presented in Fig.\,\ref{fig:resultsMWHalo}. The red solid line shows the constraints for main DM profile \cite{McMillan2017}. Other colored lines show the constraints as a function of DM profiles \cite{Eilers2019, Sofue2020, Ou2024, Cautun2020, Lim2023, Nesti2013, Lin2019}.

To account for the effect of statistical background fluctuations on the resulting constraints we generated 5000 artificial spectra for Regions I and II and repeated the line search procedure. Artificial spectra were generated from the Region II spectrum (red line on Fig.\,\ref{fig:total_signal}) using bootstrap approach. The results of these simulations are presented in Fig.\,\ref{fig:resultsMWHalo}. The black dashed line indicates the median value for the expected constraints. The green and yellow dashed areas show the confidence intervals at 1 and 2$\sigma$ significance, respectively.

\begin{table}[h]
    \centering
    \begin{tabular}{|l|c|c|c|c|}
        \hline
        Name  & \thead{RA\\(deg)} & \thead{Dec\\(deg)} & \thead{$\log_{10}(J$\\ $\times$GeV$^{-2}$ cm$^{5})$} 
        & \thead{Exposure\\(s)} \\
        \hline\hline
        And I & 11.42 & 38.04 & 16.81$^{+0.40}_{-0.36}$ & 172.0\\
        \hline
        And III & 8.89 & 36.50 & 17.03$^{+0.33}_{-0.30}$ & 185.4\\
        \hline
        And V & 17.57 & 47.63 & 17.19$^{+0.29}_{-0.25}$ & 183.2\\
        \hline
        And VII & 351.63 & 50.68 & 16.98$^{+0.16}_{-0.15}$ & 199.5\\
        \hline
        And XIV & 12.90 & 29.70 & 15.74$^{+0.38}_{-0.37}$ & 174.7\\
        \hline
        And XVIII & 0.56 & 45.08 & 16.75$^{+0.48}_{-0.44}$ & 194.1\\
        \hline
        Aquarius II & 338.48 & -9.33 & 18.27$^{+0.66}_{-0.58}$ & 67.2\\
        \hline
        Bootes I & 210.03 & 14.50 & 18.17$^{+0.31}_{-0.29}$ & 167.1\\
        \hline
        Canes Venatici I & 202.01 & 33.56 & 17.42$^{+0.17}_{-0.15}$ & 149.5\\
        \hline
        Canes Venatici II & 194.29 & 34.32 & 17.82$^{+0.47}_{-0.47}$ & 120.9\\
        \hline
        Carina & 100.40 & -50.97 & 17.83$^{+0.10}_{-0.09}$ & 263.1\\
        \hline
        Carina II & 114.11 & -57.99 & 18.25$^{+0.55}_{-0.54}$ & 280.3\\
        \hline
        Cetus & 6.55 & -11.04 & 16.28$^{+0.20}_{-0.19}$ & 97.1\\
        \hline
        Coma Berenices & 186.75 & 23.90 & 19.00$^{+0.36}_{-0.32}$ & 99.0\\
        \hline
        Crater II & 177.31 & -18.41 & 18.82$^{+0.15}_{-0.15}$ & 121.9\\
        \hline
        Draco & 260.05 & 57.92 & 18.83$^{+0.12}_{-0.12}$ & 876.3\\
        \hline
        Eridanus II & 56.09 & -43.53 & 17.28$^{+0.34}_{-0.31}$ & 296.0\\
        \hline
        Fornax & 39.99 & -34.45 & 18.09$^{+0.10}_{-0.10}$ & 217.8\\
        \hline
        Hercules & 247.76 & 12.79 & 17.37$^{+0.53}_{-0.53}$ & 179.1\\
        \hline
        Horologium I & 43.88 & -54.12 & 19.27$^{+0.77}_{-0.71}$ & 276.2\\
        \hline
        Hydrus I & 37.39 & -79.31 & 18.65$^{+0.32}_{-0.31}$ & 220.3\\
        \hline
        Leo I & 152.12 & 12.31 & 17.64$^{+0.14}_{-0.12}$ & 67.2\\
        \hline
        Leo II & 168.37 & 22.15 & 17.76$^{+0.22}_{-0.18}$ & 85.2\\
        \hline
        Leo T & 143.72 & 17.05 & 17.49$^{+0.49}_{-0.45}$ & 67.6\\
        \hline
        Reticulum II & 53.93 & -54.05 & 18.96$^{+0.38}_{-0.37}$ & 362.1\\
        \hline
        Sculptor & 15.04 & -33.71 & 18.58$^{+0.05}_{-0.05}$ & 120.5\\
        \hline
        Segue 1 & 151.77 & 16.08 & 19.12$^{+0.49}_{-0.58}$ & 67.4\\
        \hline
        Sextans & 153.26 & -1.61 & 17.73$^{+0.13}_{-0.12}$ & 68.5\\
        \hline
        Tucana II & 342.98 & -58.57 & 18.84$^{+0.55}_{-0.50}$ & 99.1\\
        \hline
        Ursa Major I & 158.72 & 51.92 & 18.26$^{+0.29}_{-0.27}$ & 87.0\\
        \hline
        Ursa Major II & 132.88 & 63.13 & 19.44$^{+0.41}_{-0.39}$ & 105.1\\
        \hline
        Ursa Minor & 227.29 & 67.22 & 18.75$^{+0.12}_{-0.12}$ & 254.9\\
        \hline
        Willman 1 & 162.34 & 51.05 & 19.53$^{+0.50}_{-0.50}$ & 87.4\\
        \hline\hline
        Ursa Major III & 174.71 & 31.08 & 21.00$^{+1.00}_{-2.00}$ & 88.5\\
        \hline
    \end{tabular}
    \caption{The dSphs galaxies used in our analysis. The total exposure time is $T = 6000$ seconds. Total area of the sky covered by the galaxies is $\Omega = 25.92$ deg$^2$. Total J-factor obeys $20.20^{+0.14}_{-0.21}=\log_{10}(J_{\text{dSph}}^{\text{tot}}/\text{GeV}^2\text{cm}^{-5}$)
    \footnote{Here $J^\text{tot}_\text{dSph} = \sum_\text{dSph} 10^{J_i}$ and $\Delta J^\text{tot}_\text{dSph} = \log{10} \sqrt{\sum_\text{dSph} ( 10^{J_i} \Delta J_i )^2 } $. The $J_i$ is the mean value for i-th dSph galaxy and $\Delta J_i$ is the largest of the error bars for i-th dSph galaxy.}.
    These values do not take into account the Ursa Major III/UNIONS 1 galaxy, which we analyze separately.}
    \label{tab:galCat}
\end{table}

\section{Spheroidal dwarf galaxies}
In the second part of our work, we focus on the similar hypothetical signal from DM annihilation in dwarf spheroidal (dSph) satellite galaxies of the MW and M31. Studies of the motion of stars in dSph galaxies show that the virial mass of such galaxies is much larger than that obtained from direct star counting. In addition, such galaxies have extremely low surface brightness. The measured mass-luminosity ratio can reach 1000 $M_\odot/L_\odot$ and even higher. Thus, the dark-to-visible mass ratios of dSph galaxies noticeably exceed that of the MW. This fact, together with the absence of bright X-ray sources there, make them promising objects for testing various models with DM candidates in the keV mass range.

\begin{figure}[bt]
    \centering
    \includegraphics[width=\columnwidth]{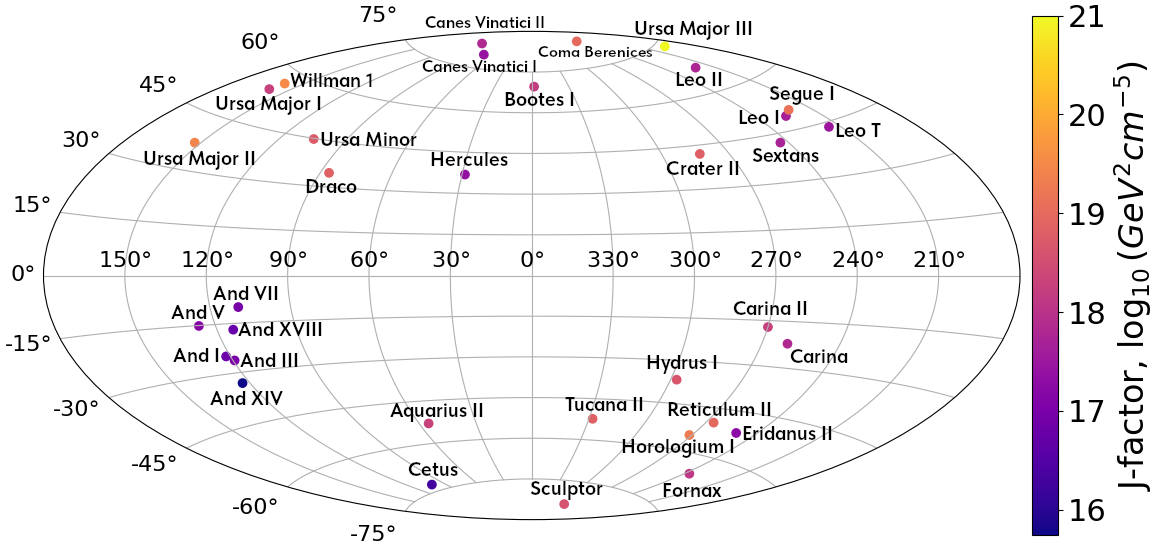}
    \caption{Distribution of satellite galaxies on the sky in the Galactic coordinate system. The color indicates the $\log_{10}(J\,/\text{GeV}^2\,\text{cm}^{-5})$ value.}
    \label{fig:dwarf_skymap}
\end{figure}

In Ref.\,\cite{galaxiesCat},  authors estimated J-factors for 41 dSph satellite galaxies of the MW and M31. We select only 31 galaxies out of 41. We exclude 10 galaxies from our analysis because either the authors gave only an upper limit on the J-factor or labeled the J-factor as ``values should be used with caution''. We also add 2 more galaxies from \cite{galaxiesCat2}: Crater II and Hydrus I. In addition,  we also take a separate look at the Ursa Major III/UNIONS 1 galaxy. This recently discovered galaxy is notable for being the ``darkest'' of all the discovered dSph satellite galaxies of the MW \cite{UM3_1, UM3_2}. A complete list of galaxies is presented in Tab.\,~\ref{tab:galCat} and Fig.\,\ref{fig:dwarf_skymap} shows the distribution of these galaxies across the sky in the galactic coordinate system.

\begin{figure}[bt]
    \centering
    \includegraphics[width=\columnwidth]{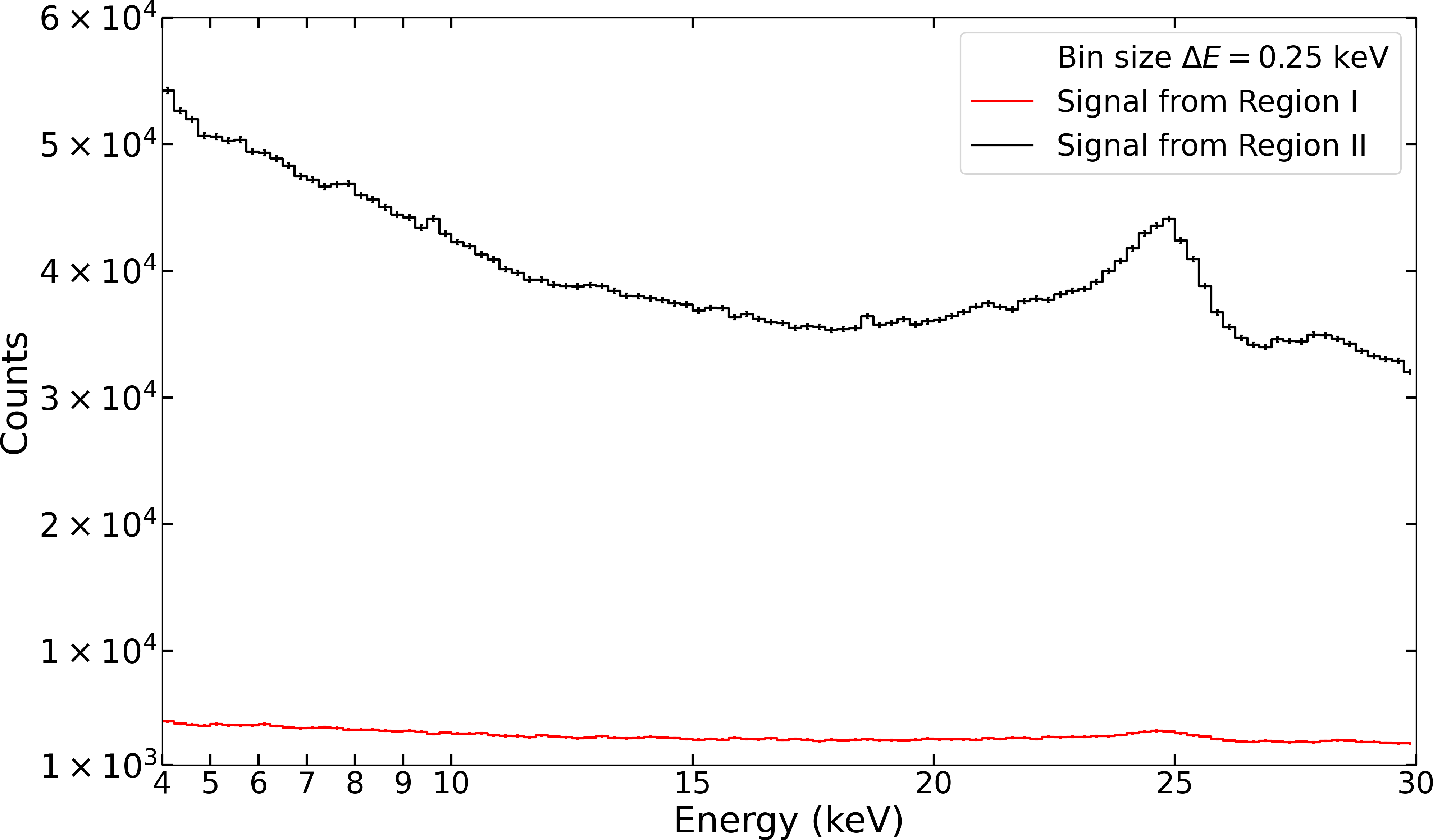}
    \caption{Raw spectrum in counts for Region I(red line) and Region II (black line) {\it summed over} all the galaxies from Tab.\,~\ref{tab:galCat} (except Ursa Major III/UNIONS 1). The cross-normalization factor $\lambda \approx 0.08$. Bin size $\Delta E = 0.25$ keV.}
    \label{fig:raw_stacked_spec}
\end{figure}

\begin{figure}[bt]
    \centering
    \includegraphics[width=\columnwidth]{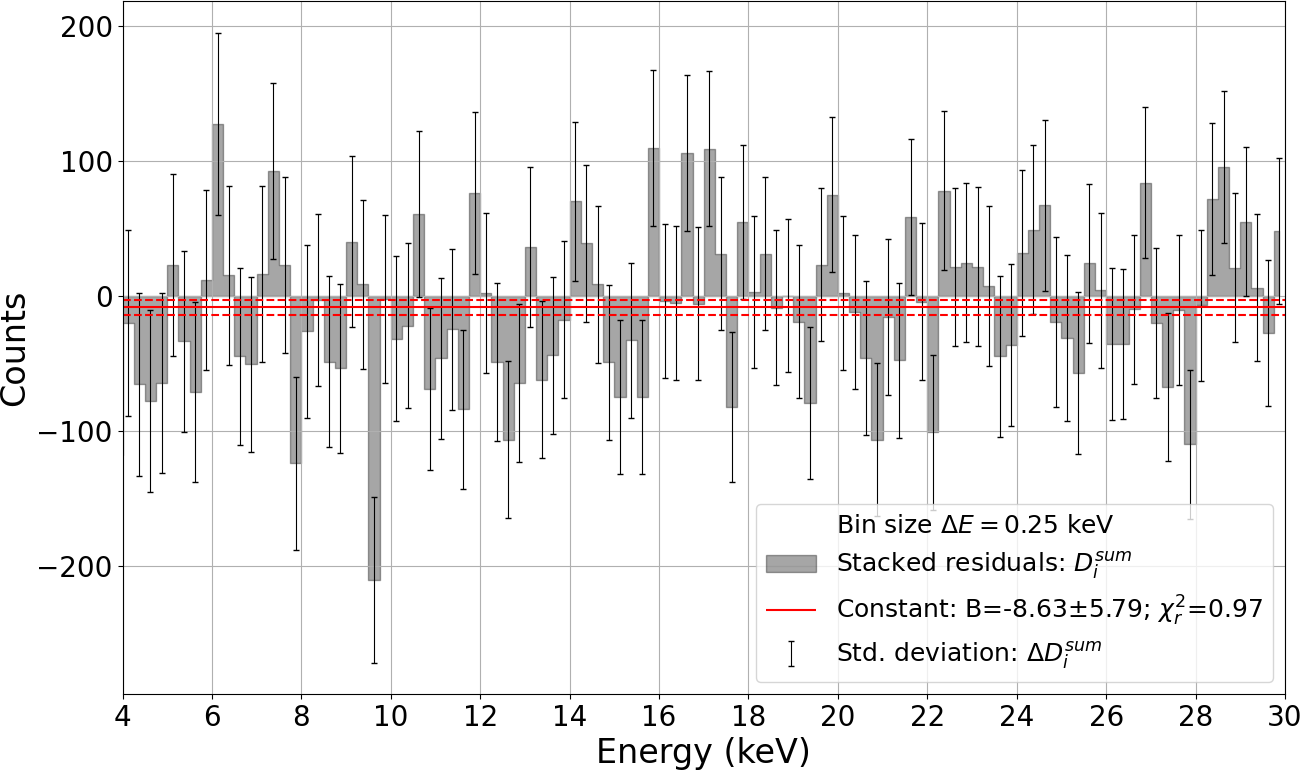}
    \caption{Residuals $D_i^{\text{sum}} \!\! = \!\! \sum_{\text{dSph}}D_i$ summed over all the galaxies from Tab.\,~\ref{tab:galCat} (except Ursa Major III/UNIONS 1). Vertical lines show the error bars $\Delta D_i^{\text{sum}} = \sqrt{\sum_{\text{dSph}} \Delta D^2_i}$. The red solid line is the fit to the residuals by constant $B$ and the red dashed lines indicate $\pm1\sigma$ confidence intervals.}
    \label{fig:stacked_spec}
\end{figure}

In the case of dSph galaxies, our method for finding constraints is very similar to the method used in the previous section. Two regions on the sky are also defined for each galaxy: Region I "Signal + background" and Region II "Background only". Region I is a circle of radius 0.5$^\circ$ around the center of dSph galaxy. This is the maximum radius for which the J-factor is calculated. Region II is a ring on the sky with angular radius from 1$^\circ$ to 2$^\circ$ centered at the source. This choice is due to the fact that the signal from the dark halo of dSph galaxy in this region will be insignificant. As in the previous section, for each galaxy we find the residuals $D_i$ as in \eqref{eq:residuals}. The cross-normalization factor $\lambda$ for each galaxy is calculated as follows:
\begin{equation}
    \lambda = \frac{\Omega^I}{\Omega^{II}}\frac{t^I}{t^{II}}.
\end{equation}
Here $\Omega^I=0.785$ deg$^2$, $\Omega^{II}=9.425$ deg$^2$ and $t^I$, $t^{II}$ are the area and exposure time for Regions I and II respectively. Since the regions around dwarf galaxies are small and surveyed almost uniformly, for every galaxy the ratio $t^I/t^{II}$ is almost one. Fig.\,\ref{fig:raw_stacked_spec} shows the raw spectra $N^{I}_i$ and $N^{II}_i$ and Fig.\,\ref{fig:stacked_spec} shows the residuals $D_i$ {\it summed over} all the galaxies (except Ursa Major III/UNIONS 1, see below).

When calculating the expected signal from the annihilation of DM particles, we also need to take into account the fact that we look at dark halos of satellite galaxies through the MW Halo. Expected signal from Region I and II:
\begin{align}
    F^I = \frac{\sigma v}{4 \pi m^2_\chi}& \times \l J_{\text{dSph}} + J_{\text{MW}\rightarrow\text{dSph}}(0^\circ, 0.5^\circ) \r\\
    F^{II} \approx \frac{\sigma v}{4 \pi m^2_\chi}& \times \l J_{\text{MW}\rightarrow\text{dSph}}(1^\circ, 2^\circ) \r,
\end{align}
where $J_{\text{MW}\rightarrow\text{dSph}}(\theta_1, \theta_2)$ is the J-factor for MW in a given cone toward the dSph galaxy. To find this value, we used the J-factor distribution map (Fig.\,~\ref{fig:DM}) provided in HEALPix format. The MW J-factor is the sum of pixels in the corresponding region multiplied by the size of one pixel (we used $N_\text{side}$ = 1024 and the pixel size is $4\pi/(12N_{\text{side}}^2) = 10^{-6}$ sr ).

After subtraction, given the fact that $J^{\text{add}}_{\text{MW}} = J_{\text{MW}}(0^\circ, 0.5^\circ) -\lambda J_{\text{MW}}(1^\circ, 2^\circ) \approx 0$, we get the following expression for the expected signal:
\begin{equation}
    F = F^I - \lambda F^{II} \approx \frac{\sigma v}{4 \pi m^2_\chi} \times \l J_{\text{dSph}} + J^{\text{add}}_{\text{MW}} \r.
\end{equation}
Here $J^{\text{add}}_{\text{MW}}$ refers to the contribution from the MW. For the MW satellites, $J^{\text{add}}_{\text{MW}}/J_{\text{dSph}} \ll 0.1$. However, for some M31 satellites, the relative contribution from the MW can be $\sim 0.2$ and even more \footnote{Here we do not consider the contribution from M31 and LMC due to their insignificance. For M31 with the NFW profile from \cite{M31Halo}, we obtain $J_{M31}/J_{MW}<0.01$ or less. For LMC with the Hernquist profile from \cite{LMCHalo}, we obtain $J_{LMC}/J_{MW} < 0.025$ or less.}.

\begin{figure}[tb]
    \centering
    \subfigure{\includegraphics[scale=0.25]{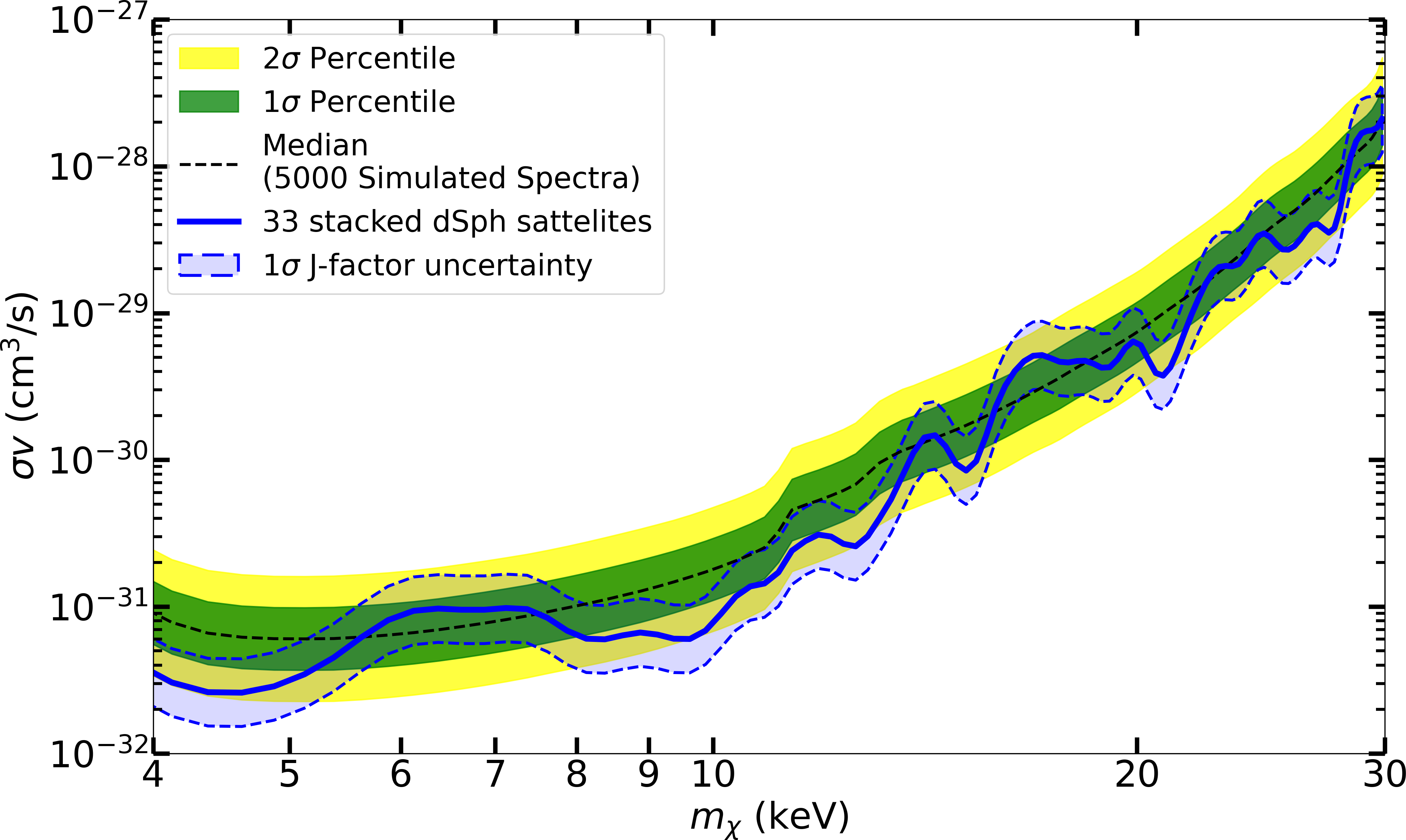}}
    \subfigure{\includegraphics[scale=0.25]{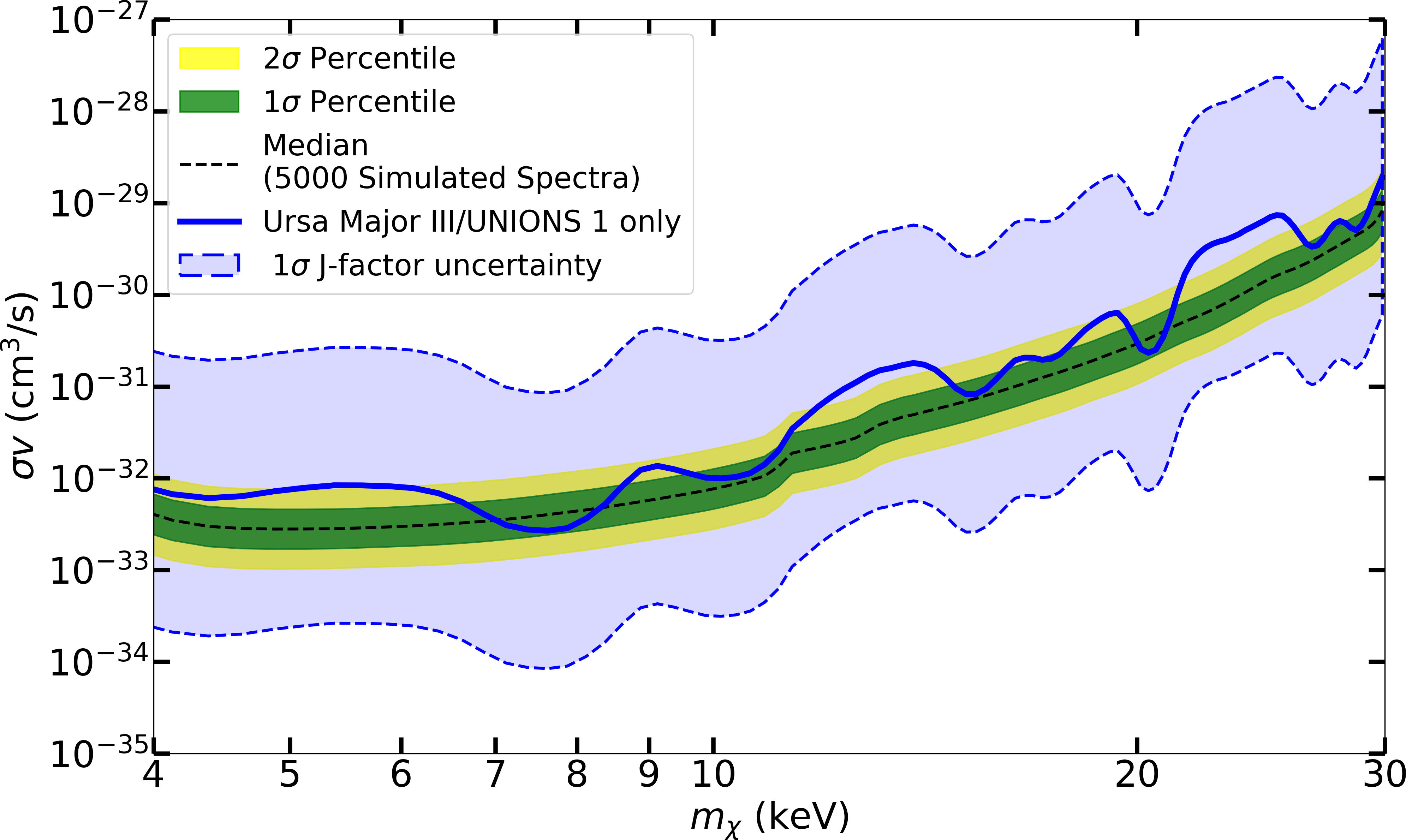}}
    \caption{Upper limits on s-wave annihilation cross section (95\% C.L.) derived from observations of Local Group dSph satellite galaxies. \textit{Top panel} is for 33 dSph galaxies and \textit{bottom panel} is for Ursa Major III/UNIONS 1 galaxy only. The black dashed lines show the median values for the expected limits obtained with analysis of 5000 artificial spectra of background. The green and yellow areas correspond to the 1$\sigma$ and 2$\sigma$ percentiles, respectively. Blue shaded areas with dashed boundaries show 1$\sigma$ errors associated with the uncertainty in the J-factor estimate.}
    \label{fig:results_dSph}
\end{figure}

The further procedure for finding the signal and the upper 95\% limit on it is completely similar to the procedure used in the previous Section. Fig.\,\ref{fig:raw_stacked_spec} shows that each bin contains thousands of events in Region I and tens of thousands in Region II. This fact again allows us to use the log-likelihood function \ref{eq:LogLike}. However, the relation between the number of registered photons $A$, the velocity-independent cross section $\sigma v$ and particle mass $m_\chi$ (which equals the energy of the registered photon,  $m_\chi = E$) is expressed as follows:
\begin{equation}
    \sigma v = \frac{4 \pi m^2_\chi A \Omega}{T G_8(E) \l J^{\text{tot}}_{\text{dSph}} + J^{\text{add tot}}_{\text{MW}} \r}.
\end{equation}
Here $T$, $\Omega$ and $J^{\text{tot}}_{\text{dSph}}$ are the total exposure time, area, and J-factor, respectively,  see Tab.\,~\ref{tab:galCat}. The $G_8(E)$ is the  grasp function, see Fig.\,\ref{fig:grasp}. The area under line $A$ is a 95\% upper limit on the number of photons associated with annihilation of DM particles (see \ref{eq:95cl}). And $J^{\text{add tot}}_{\text{MW}}$ is the total contribution from MW. Note that $J^{\text{add tot}}_{\text{MW}}/J^{\text{tot}}_{\text{dSph}} = 6\cdot10^{-5}$. The resulting constraints are summarized in top panel of Fig.\,\ref{fig:results_dSph}.

We pay a special attention to the Ursa Major III/UNIONS 1 galaxy. It follows from \cite{UM3_1} that the J-factor for this galaxy is estimated to satisfy $21^{+1}_{-2}=\log_{10}(J\,/\text{GeV}^2\,\text{cm}^{-5}$). Even taking into account the large uncertainty in the definition of the J-factor, this fact makes this galaxy the darkest of all known dSph galaxies. The expected signal from this galaxy alone is $J_{\text{UMa III}}/J^{\text{tot}}_{\text{dSph}} = 6.3$ times larger than from all dSph satellite galaxies combined. Therefore, we do not include this galaxy in the general list and consider it separately. The results of this analysis are presented in bottom panel of Fig.\,\ref{fig:results_dSph}. The obtained limit is weaker than those from the MW, but stronger than that from the other dwarf galaxies. Note that recent analysis \cite{Zhao:2024say} reveals unhealthy dependence of the estimated $J$-factor on the star velocity measurements: exclusion of the single largest velocity outlier extremely reduces $J$-factor and hence weakens the associated with it limits on the DM cross section.

\section{Conclusion}
\begin{figure}[tb]
    \centering
    \includegraphics[width=\columnwidth]{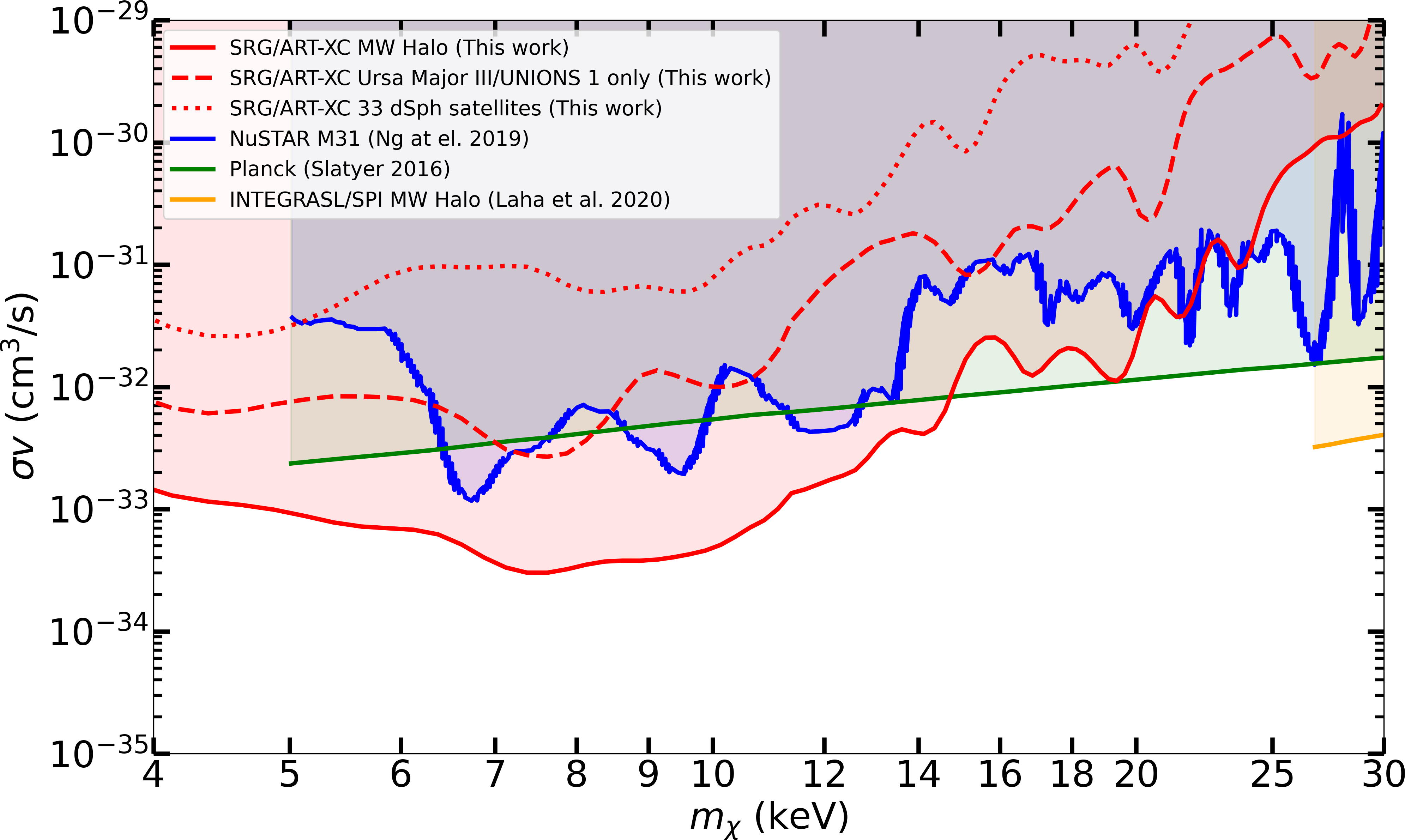}
    \caption{Upper limits (red solid line) on s-wave annihilation cross section (95\% C.L.) from analysis of SRG/ART-XC observation of MW Halo (for NFW profile \cite{McMillan2017}). The red dotted and red dashed lines shows the constraints obtained from observations of 33 dSph galaxies of the Local Group and separately for Ursa Major III/UNIONS 1 dSph galaxy, respectively. The blue solid line show limits obtained with the NuSTAR in the M31 \cite{nustarM31}. The green solid line show limits obtained with Planck data \cite{Planck}. And the orange solid line show limits obtained with INTEGRAL/SPI in the MW halo \cite{Laha:2020ivk}.}
    \label{fig:results_final}
\end{figure}

In this work we perform the analysis of SRG/ART-XC data looking at possible peaklike signature in diffuse X-ray spectrum associated with DM annihilation. 
To summarize, we find that the  SRG/ART-XC data collected in the survey mode over 4 full-sky surveys, naturally make MW as the most promising source of the DM annihilation to be searched for. The obtained from MW limit is strongest among all we found in literature, for the DM mass range $m_\chi=4-15$\,keV, and is competitive to those from NuSTAR observation of M31 galaxy and from analysis of the Planck data for $m_\chi=15-25$\,keV, as shown in Fig.\,\ref{fig:results_final}. 
It is worth to note that at the lower range of mass this limit may further strengthened with similar analysis of the all-sky survey data of the second SRG telescope, SRG/eROSITA.

\section{Acknowledgements}
The work is supported by the RSF Grant No. 22-12-00271.

\appendix
\section{The line search method validity}
The main point of our line search algorithm is to maximize the likelihood function sequentially at each point of the spectrum. The maximization is performed on the only one free parameter, which is the area $A$ under the line (which corresponds to the number of photons forming the line). Only $A>0$ is allowed. In this case, we calculate the significance of the signal as:
\begin{equation}
    S = \sqrt{2}[l(D|A=0,E) - l(D|A=\hat{A},E)]^{1/2}
\end{equation}
Here $l(D|A,E)$ is the log-likelihood function. Both in the case of MW Halo and dSph galaxies we use a likelihood function of the form \ref{eq:LogLike}.

\begin{figure}[t]
    \centering
    \includegraphics[width=\columnwidth]{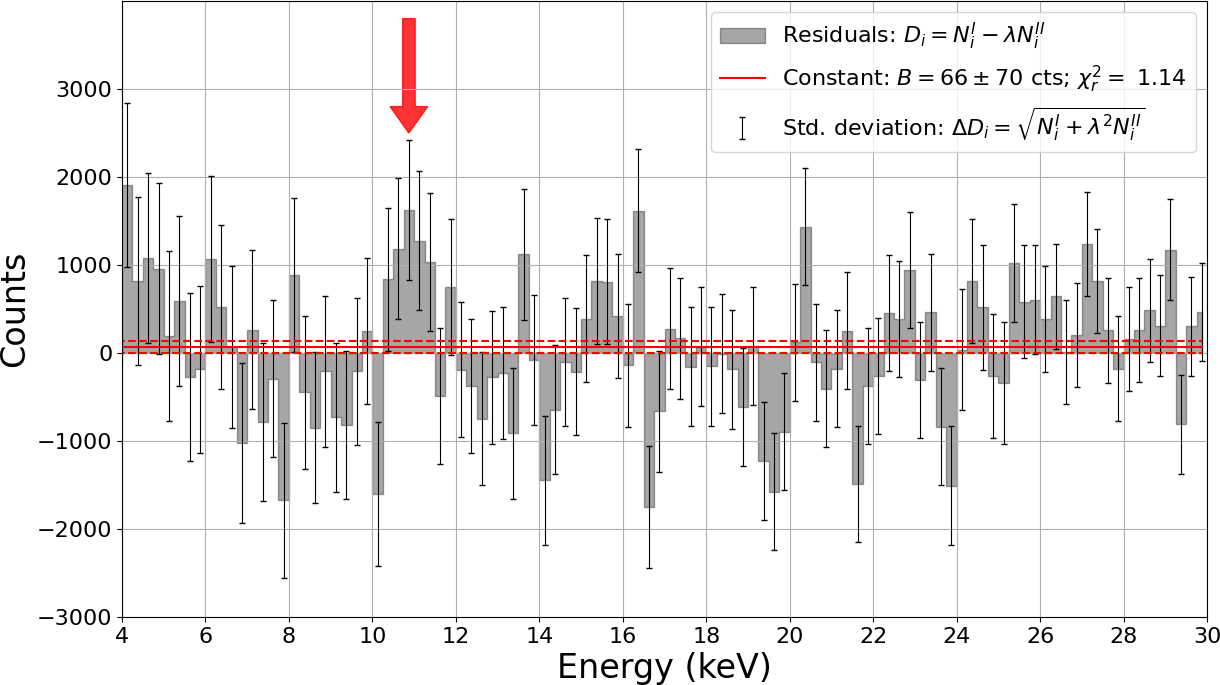}
    \caption{After subtraction residuals of two spectra with artificial line at 11 keV (under the red arrow).}
    \label{fig:residuals_withLine}
\end{figure}

To demonstrate the validity of the line search method, we can insert a line with known position and area into the original MW Halo data and test our algorithm on this data. As an example, we added $10^4$ photons from the annihilation of a particle with mass 11 keV to the spectrum of Region I. In the spectrum of Region II, this line is also at 11 keV, but its area is 3 times smaller (because $J^I/J^{II} \approx 3$). After subtraction, we expect to detect at 11 keV a line with area $A^I - \lambda A^{II} \approx 9150$. Fig.\,\ref{fig:residuals_withLine} shows the residuals after subtraction procedure. This figure is completely identical to Fig.\,\ref{fig:residuals} with the only exception that we now clearly see the line at 11 keV.

\begin{figure}
    \centering
    \subfigure{\includegraphics[width=\columnwidth]{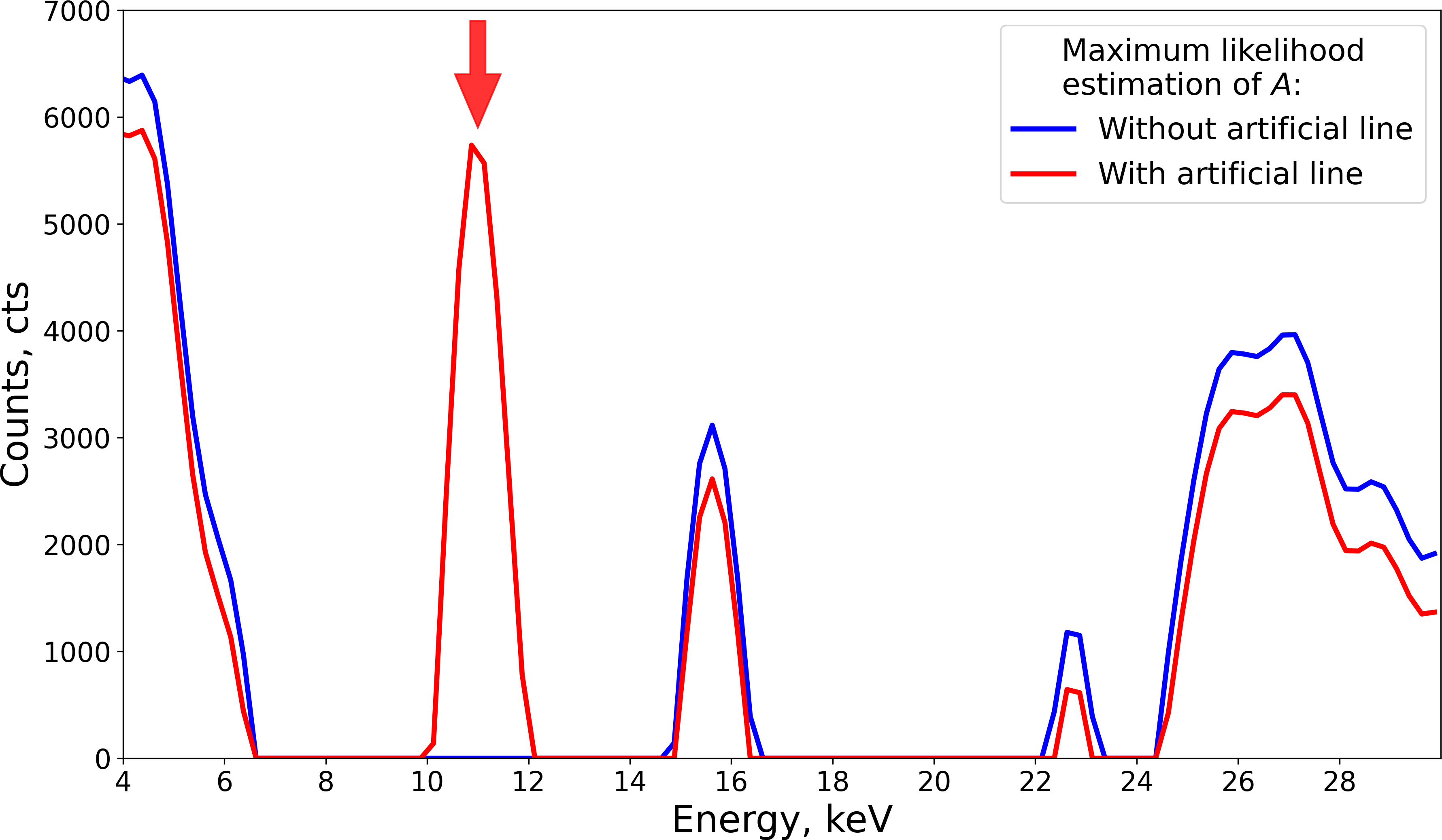}}
    \subfigure{\includegraphics[width=\columnwidth]{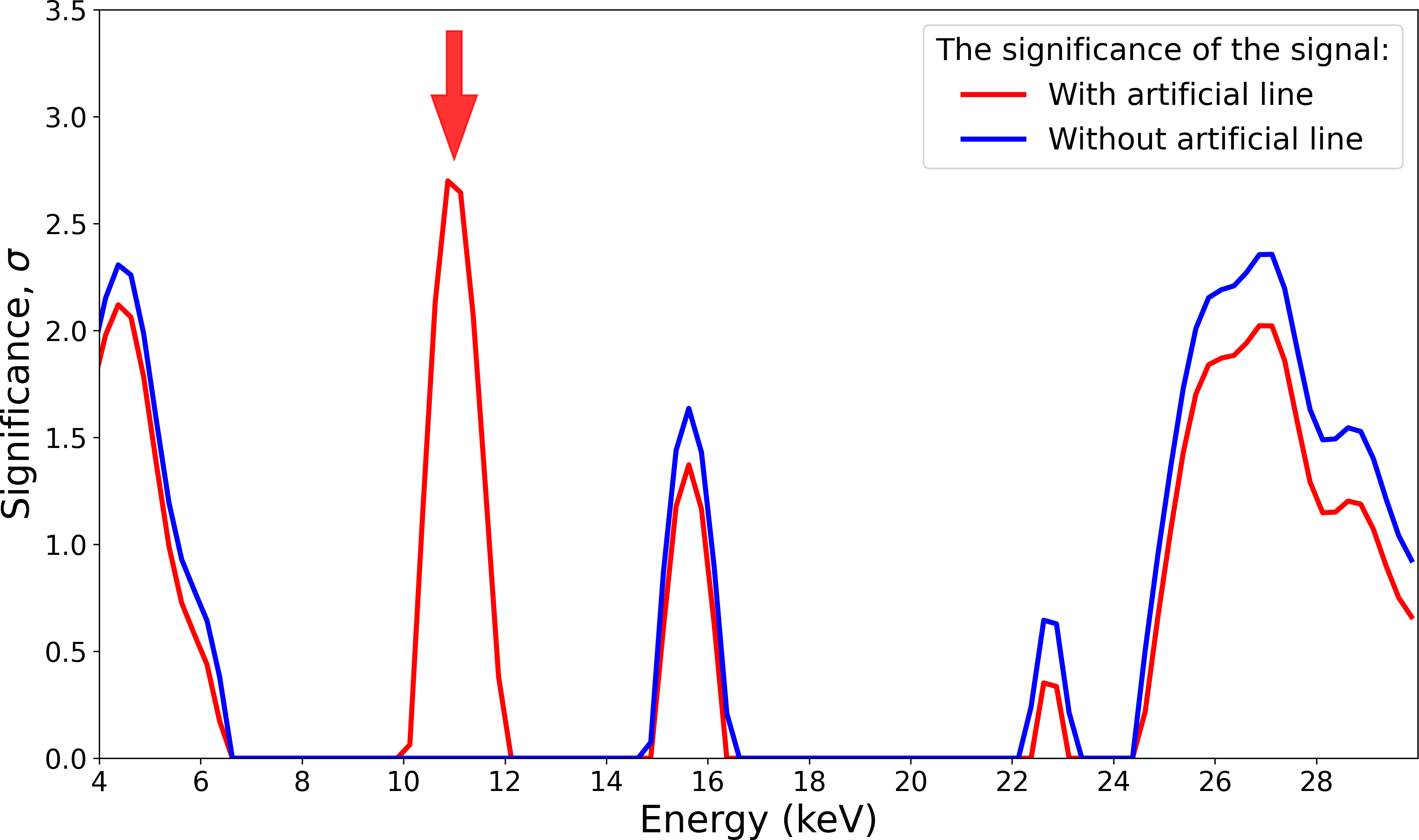}}
    \caption{The \textit{top panel} shows the maximum likelihood estimate of A at each point in the spectrum with (red solid line) and without (blue solid line) the artificial line. The \textit{bottom panel} shows the significance of the A estimates at each point in the spectrum with (red solid line) the artificial line and without (blue solid line). The red arrow shows the original position of the artificial line.}
    \label{fig:ampl_sig}
\end{figure}

\begin{figure}
    \centering
    \includegraphics[width=\columnwidth]{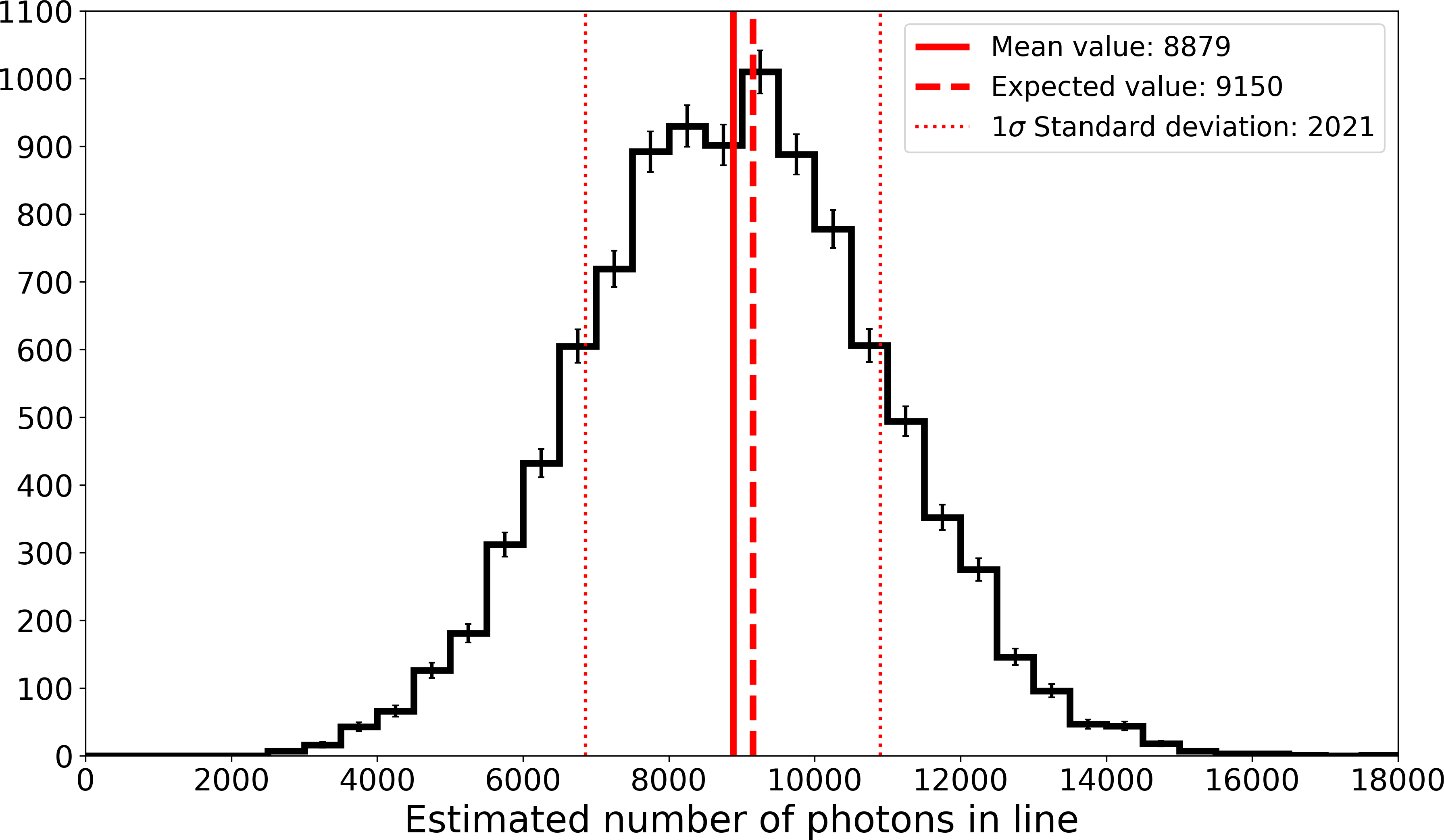}
    \caption{Distribution of the maximum likelihood estimate of $A$ after $10^4$ iterations of the line search algorithm over a random background. The red dashed line shows the original value of $A$. The red solid line shows the measured value of $A$ and dotted red line shows its 1$\sigma$ confidence interval.}
    \label{fig:A_distribution}
\end{figure}

Fig.\,\ref{fig:ampl_sig} shows that our line search algorithm also indicates that a significant line has appeared at 11 keV. The change in significance everywhere in the spectrum is due to a change of the residual constant $B$. However, the maximum likelihood estimate of A is 5736, that is different from the expected value. This is due to the realization of a Poisson (in case of millions events - Gaussian) background under the line. Using bootstrapping and generating a background below the line, we can show that our algorithm correctly determines the area of the line. Fig.\,\ref{fig:A_distribution} shows the results of such a procedure. After $10^4$ iterations, the average measured number of photons in the line is very close to the expected value.

Thus we see that this annihilation line search algorithm is able to correctly estimate the position and area of the lines. And deviations in the estimation of the line area are related to the realization of the background under the line and can be accounted for by Monte Carlo simulations.

\addcontentsline{toc}{chapter}{\bibname}
\bibliographystyle{apsrev4-1}
\bibliography{bibliography}

\end{document}